# SINGLE-SPIN MICROSCOPE WITH SUB-NANOSCALE RESOLUTION BASED ON OPTICALLY DETECTED MAGNETIC RESONANCE

Gennady P. Berman and Boris M. Chernobrod

*Theoretical Division,T-4 and CNLS, Los Alamos National Laboratory, Los Alamos, New Mexico 87545*

ABSTRACT

We summarize our new scanning magnetic 3-D imaging system. This scanning system uses optically detected magnetic resonance in a single nitrogen vacancy center in a diamond nanocrystal. The theoretical analysis and the first experimental demonstrations have proved that this method has single spin sensitivity and a sub-nanoscale spatial resolution at room temperature.

The invention of the scanning tunneling microscope (STM) and the atomic force microscope (AFM) initiated a new era of material science and technology characterized by 2-D imaging with atomic resolution and manipulation of individual atoms. For further progress in material science, and especially in structural biology, 3-D imaging with sub-nanometer resolution is very desirable. Currently the most promising technique for 3-D imaging is magnetic resonance force microscopy (MRFM), which senses individual electron spins [1,2] with nanoscale resolution and can detect the collective magnetization of about 100 nuclear spins [3]. The highest sensitivity demonstrated by MRFM uses a time modulation technique called the oscillating cantilever-driven adiabatic reversals (OSCAR) which requires a long phase relaxation time, $T_2$, of measured spins, which usually requires rather low temperatures. For example, a temperature of 300 mK was used for 3D imaging of the tobacco mosaic virus [3]. This limitation is incompatible with the room-temperature operation needed for the study of biological systems under physiological conditions.

Recently we proposed and patented an alternative approach that potentially has single spin sensitivity, sub-nanometer spatial resolution, and the ability to operate at room temperature [4-6]. In our approach, a nanoscale photoluminescent center (PC) exhibits an optically detected magnetic resonance (ODMR) in the vicinity of a magnetic moment in the sample related to unpaired individual electron spins or nuclear spins, or an ensemble of spins.
It is well-known that it is possible to detect a single molecule using conventional ODMR techniques. The lateral resolution of the conventional ODMR method is limited by the size of light spot. The highest resolution, 30 – 50 nm, is obtained using a near-field scanning optical microscope (NSOM). Another limitation of the conventional ODMR technique is that the unpaired electron has to be a part of the molecule that absorbs or emits light. We proposed a modification of the ODMR technique which is free from these limitations. We proposed several scanning strategies [4,5]. One of them is shown in Figure 1, where a photoluminescent center is located at the tip of an AFM cantilever. Laser radiation excites luminescence of the PC, and a nearby radiofrequency (RF) coil produces an oscillating field at the frequency resonant to the transition between the magnetic sublevels of the PC. (See Figure 2.) Population redistribution of magnetic sublevels causes a decrease in the luminescence intensity. The dependence of luminescence intensity as a function of the RF frequency or amplitude of external static magnetic field exhibits a narrow resonance. The EPR frequency shift due to the external magnetic field is the subject of measurements. We considered two types of promising sensor materials that exhibit ODMR properties [6]: 1) nitrogen-vacancy (N-V) centers in diamond and 2) CdSe nanoparticles. Although each of these materials is optically active and has reasonably sharp photoluminescence peaks, N-V centers in diamond have the significant advantages of having extraordinary chemical and photostability, very long spin lifetimes, and single-spin detection capability at room temperatures.

A comprehensive theoretical analyses of N-V centers in diamond as a promising candidate sensor for our method was published recently by Degen (IBM Almaden Research Center) [7].

He considered two types of measurements, continuous and pulsed RF waves. For continuous RF waves, the amplitude of the RF field is constant in time, and the sensitivity of proposed method is determined by the smallest measurable EPR frequency shift, which is comparable with the EPR homogeneous line width. In this case the minimal resolvable magnetic field is approximately 0.2 mT, and a single electron spin can be detected at distance of $r \leq 3nm$ from the N-V center. The corresponding 3D spatial resolution is of the same order of magnitude. For detecting a single proton spin, the distance is $r \leq 0.3nm$. However, by observing the spin precession in spin echo-type experiments it is possible to obtain much better sensitivity. (Note that the main factors in the phase relaxation of N-V centre spin are its interactions with the $C^{13}$ nuclear spins and with the electronic spin impurities. The phase memory time for ultrapure diamond approaches one millisecond.) In the case of using phase locking of the detection system to the oscillatory motion of the cantilever, and using echo-type techniques, where the RF field amplitude is modulated in time with pulse durations shorter than $T_2$, the minimal measurable magnetic field is approximately 60 nT, and the distance for detection of single electron spin is $r \leq 40nm$. For a single proton spin the corresponding distance is $r \leq 5nm$.

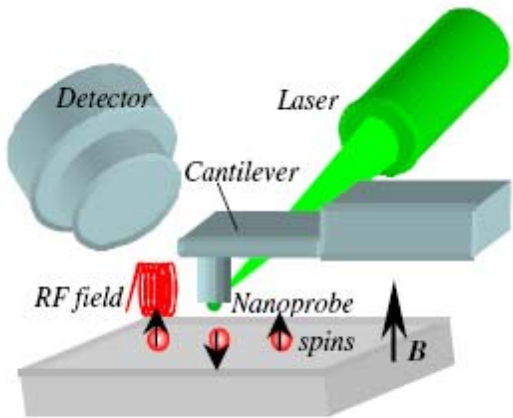

Figure 1. AFM-ODMR setup.

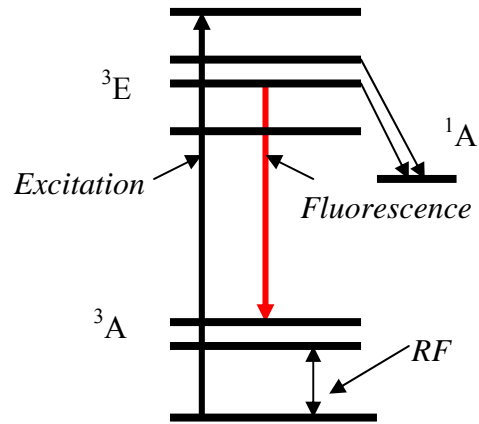

Figure 2. Transitions in the N-V center.

Quite recently our proposed technique was implemented by a collaborating group of scientists from Germany and the US [8]. Our nanoscale scanning technique was realized using the single spin of N-V center in a diamond nanoparticle. This group performed two types of experiments. In the first experiment, the magnetic tip of an AFM was scanned at a known distance above the diamond nanocrystal, and magnetic field experienced by the single N-V center was recorded using ODMR. The ultimate resolution was about 5 nm. Because the N-V center was localized to a fraction of a nanometer in the diamond lattice, the resolution scale was smaller than the size of the magnetic tip and the nanocrystal. In their second type of experiment, a nanocrystal containing a single N-V center was attached to the tip of an AFM cantilever. This scanning magnetometer was used to profile the magnetic field produced by a nanometer-size magnetic structure. The microwave frequency was in resonance with the transition between magnetic sublevels of the N-V center at the tip of the cantilever. Their measured spatial resolution was 20 nm. The corresponding magnetic resolution was 0.5 mT. The authors noted that resolution was spatially limited by the oscillatory motion of the nanocrystal attached to the AFM tip. The authors stated that the resolution could be significantly improved by phase locking the detection system to the oscillatory motion of the cantilever, and using echo-type techniques with the echo period matched to a single oscillatory period of the cantilever. The measurement accuracy could then be improved by a factor of 150 (3 μT) with corresponding sub-nanometer spatial resolution.

Another group (Harvard and MIT, US) [9] has demonstrated the very high sensitivity of a single N-V center ODMR to weak magnetic fields using a spin-echo technique. Their experiments with a single N-V center in commercially available diamond nanocrystals demonstrated a sensitivity of 0.5 μT Hz$^{1/2}$. The authors proposed to improve this sensitivity by using isotopically pure diamond with low concentrations of both $C^{13}$ and nitrogen electron impurities, in which much longer coherence and interrogation time should be possible.

The feasibility demonstrations of scanning technology and spin-echo techniques with single spin resolution constitute an emergence of a novel magnetometer with potentially single-electron and nuclear-spin sensitivity, sub-nanometer resolution volumetric imaging, and the capability to operate at room temperatures. This magnetic scanning microscope is now expected to have many broad applications from 3D *in situ* imaging of biological structures to quantum computing.